\documentstyle[preprint,aps]{revtex}
\tighten
\draft
\widetext
\input{epsf.sty}
\preprint{CLNS 01/1734}
\font\blackboard=msbm10 
\font\blackboards=msbm7 \font\blackboardss=msbm5
\newfam\black \textfont\black=\blackboard
\scriptfont\black=\blackboards \scriptscriptfont\black=\blackboardss
\def\Bbb#1{{\fam\black\relax#1}}

\def\Le{\Lambda_{eff}}

\def\ki{k_i}
\def\yi{y_i}

\def\be{\begin{equation}}
\def\ee{\end{equation}}
\def\baray{\begin{eqnarray}}
\def\earay{\end{eqnarray}}

\begin{document}
\title{Particle Horizon and Warped Phenomenology}

\medskip

\author{Horace Stoica, S.-H. Henry Tye and Ira Wasserman}

\medskip

\address{Laboratory for Nuclear Studies and Center for Radiophysics
and Space Research \\
Cornell University \\
Ithaca, NY 14853}
\medskip
\date{\today}
\maketitle

\begin{abstract}

Giant resonances of gravity Kaluza-Klein modes (with tensor couplings)
in high energy collisions are expected in the Randall-Sundrum orbifold 
model that incorporates a plausible solution to the hierarchy problem. 
When the model is extended to incorporate an exponentially small 
4-D cosmological constant, the KK spectrum becomes continuous,
even in the compactified case.
This is due to the presence of a particle horizon, which provides a 
way to evade Weinberg's argument of the need of fine-tuning to get a 
very small cosmological constant.

\end{abstract}

\section{Introduction}

        Recently Randall and Sundrum used warped geometry to propose 
a plausible solution to the hierarchy problem \cite{RS2}. In their 
$S^1/\Bbb{Z}_2$ orbifold model, the two branes sit at the two fixed end 
points. As a result of the compactification, the discrete gravity 
Kaluza-Klein (KK) modes have relatively strong (compared to the graviton) 
couplings to matter fields on the visible brane. This implies giant 
resonances in high energy collisions\cite{RS2,RS1}.

The warped geometry idea was recently used to provide a plausible 
solution to the cosmolgical constant problem\cite{ira}. More recently, 
in Ref\cite{fjstw}, this scenario was further extended to incorporate 
the hierarchy solution of the original proposal\cite{RS2}. In this paper,
we point out that the phenomenology changes drastically 
in this scenario, even though the hierarchy solution in this new scenario
is very similar to that of the $S^1/\Bbb{Z}_2$ orbifold model \cite{RS2}. 
Furthermore, this change, from a discrete to a continuous KK spectrum, 
happens irrespective
of whether the extra dimension is compactified or not. It has to do with 
the presence of the particle horizon that invariably appears in any 
scenario of the type of Ref\cite{ira,fjstw} that naturally incorporates 
an exponentially small 4-D cosmological constant. In some sense, the 
phenomenology of the continuous gravity KK spectrum in this scenario 
(compactified or not) is quite similar to that on the probe 
brane \cite{lykkenR} in the presence of the Planck brane in the 
uncompactified model also proposed by Randall and Sundrum \cite{RS1}. 

To be explicit, let us first discuss the specific two brane 
compactified model presented in Ref\cite{fjstw}. 
(We shall discuss the general situation in a moment.) 
Consider two parallel $3$-branes with brane tensions $\sigma_0>\sigma_1>0$
sitting in a compactified $5$th (i.e., $y$) dimension, with 
circumference $L_2-L_0$.
The $\sigma_0$ (Planck, hidden) brane sits at $L_0=0$ and the $\sigma_1$ 
(TeV, visible) brane sits at $L_1$. All matter fields of the standard 
model of strong and electroweak interactions are confined on the 
visible brane.
Since $L_2$ is identified with $L_0=0$, this means the branes
are separated by $L_1$ on one side and by $L_2-L_1$ on the other side.
Without loss of generality, let $L_2-L_1>L_1$. 
Using the metric ansatz
\be\label{metric}
ds^2=dy^2+A(y)[-dt^2+\exp(2Ht)\delta_{ij}dx^idx^j].
\ee
where, in the absence of matter density, the Hubble constant $H$ is 
truly a constant. 
The Einstein equation yields the general solution \cite{kaloper}
\be
\label{metricA}
  A(y) = {H^2\sinh^2[\ki(y-\yi)]\over\ki^2},
\ee
in which $\ki$ ($k_i^2=\kappa^2|\Lambda_i|/6$ where $\kappa^2$ is the 
5-D gravitational coupling and $\Lambda_i$ are the bulk cosmological 
constants) and $\yi$ are integration constants to 
be fixed by the boundary conditions at the branes.
The warp factor $A(y)$ is schematically shown in Figure 1. 

\medskip

\begin{center}
  \epsfbox{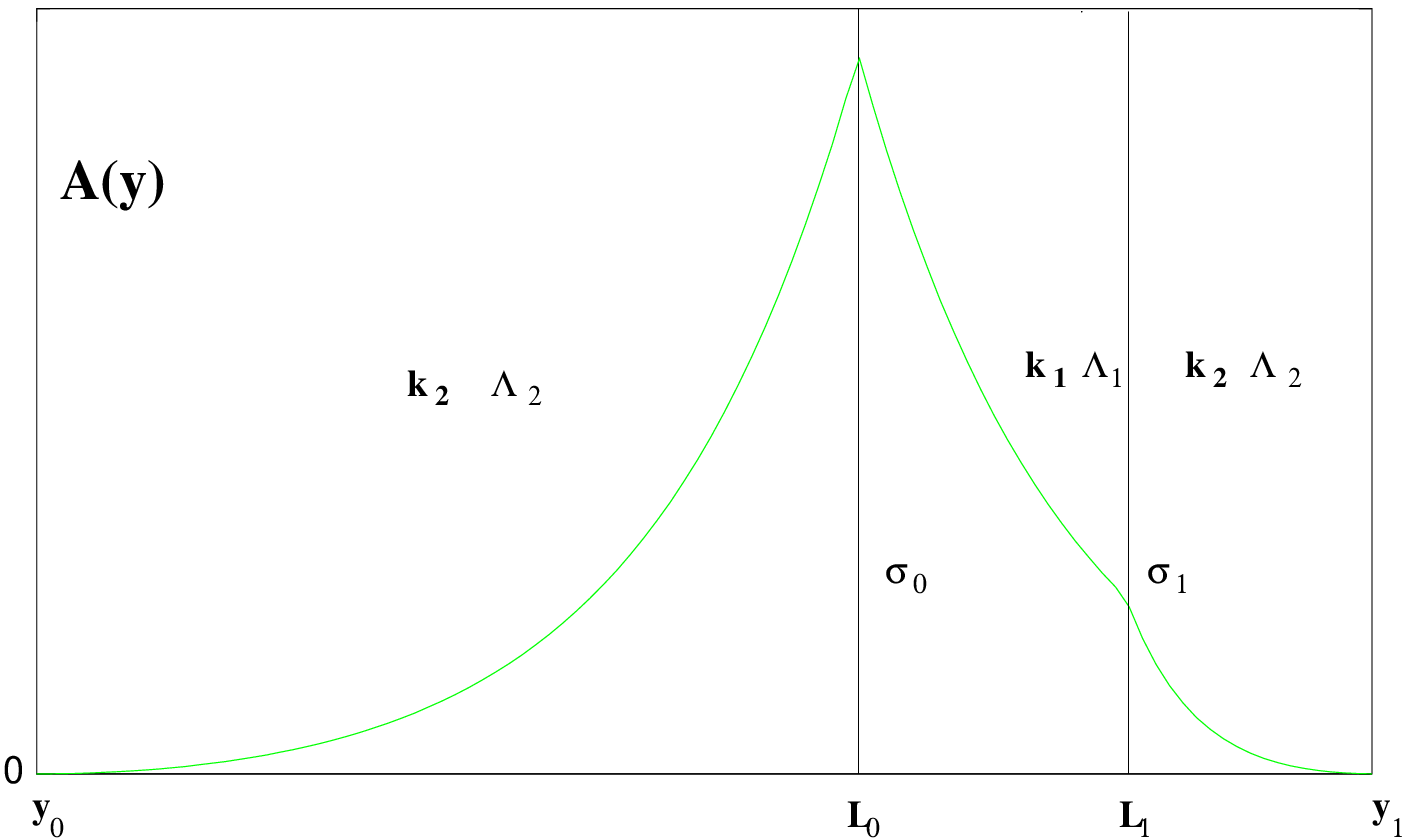}
  \parbox{12cm}{\vspace{.4cm}
    FIG.1 \hspace{2pt} The two brane compactified model, where
$y_1$ is identified with $y_0$, the position of the particle horizon.
The circle has circumference $L_2-L_0$. 
The brane at $L_0$ ($L_1$) is the Planck (visible) brane. 
The warp factor $A(y)$ is shown schematically.\vspace{12pt}}
\end{center}

We find that, for large brane separation $L_1$, the warp factor 
$A(y)$ provides a plausible explanation of the hierarchy problem: why 
the electroweak scale $m_{EW}$ is so much smaller than the Planck scale
$M_{Planck}$,
\be
\label{warpfactor}
{m_{EW}^2\over M_{Planck}^2} \simeq{A(L_1)\over A(0)}
\simeq e^{-\kappa^2(\sigma_0-\sigma_1)L_1/3}
\ee
In this case, the effective 4-D cosmological constant $\Le$ is also 
exponentially small,
\be
\label{Lecc}
\Le \simeq \frac{2\sigma_0(\sigma_0 + \sigma_1)}
{(\sigma_0-\sigma_1)}
e^{-\kappa^2[(\sigma_0+\sigma_1)(L_2-L_1)+(\sigma_0-\sigma_1)L_1]/6}.
\ee
where $H^2= \kappa_N^2 \Le/3$, and the 4-D gravitational coupling 
$\kappa_N^2 = 8 \pi G_N = 8 \pi M_{Planck}^2$ is given by
\be
\label{GNdefine}
 \frac{1}{2\kappa_N^2} = \frac{1}{2\kappa^2} \int A(y) dy,
\ee
With the above warp factor (\ref{warpfactor}) and $\Le$,
the hierarchy problem and the cosmological constant problem may be
simultaneously solved for appropriately large brane separations. 

Note that the warp factor (\ref{warpfactor}) is very similar to that 
in the Randall-Sundrum (RS) model\cite{RS2}. In fact, it reduces to 
theirs if $\sigma_1$ is taken to be negative and $\sigma_1 = -\sigma_0$,
in which case, we see that $\Le$ and $H$ become zero independent of 
the brane separations, another property of the RS model. 
Although it is consistent to have a negative tension 
brane sitting at an orbifold fixed point if its fluctuating modes 
are removed by the orbifold projection, the visible brane in our model 
is not sitting at an orbifold fixed point, so stability requires its 
brane tension to be positive. Besides the above fine-tuning
$\sigma_1 = -\sigma_0$, the RS model also requires a fine-tuning 
between the brane tension and the bulk cosmological constant. 
In Ref\cite{ira,fjstw}, the bulk cosmological constants are treated as 
integration constants which are determined by the 5-D Einstein equation.
This is made possible with the introduction of $5$-form field strengths 
and/or unimodular gravity.

We see that the metric factor $A(y_0)=0$ at $y_0$ somewhere between the 
two branes. This is a particle horizon. 
The presence of the particle horizon is quite generic in this scenario
for a non-zero $H$, 
independent of whether the 5th dimension is compactified or not.
It is also present in scenarios with more than two branes. The particle 
horizon persists even when the branes are moving slowly. 
This feature is actually highly desirable because it evades Weinberg's 
argument\cite{weinberg} on the need of fine-tuning to get a very 
small cosmological constant. For our purpose,
let us rephrase Weinberg's argument, which roughly goes as follows. 
Starting from any higher dimensional theory with 
the extra dimensions compactified, integrate out the extra dimensions
to obtain an effective low energy 4-dimensional theory. 
Suppose the 4-D cosmological 
constant is classically exponentially small, or zero. Then 4-D quantum 
effects will introduce corrections to the 4-D cosmological 
constant that will not be exponentially small. This implies 
that an exponentially small (or zero) 4-D cosmological constant in our 
universe must be the result of a fine-tuning.   

Now let us see how the warped geometry and the particle horizon evades 
this argument. For an observer on the visible brane at $y=L_1$, it 
takes infinite time for a light-like signal to travel from the brane 
to the particle horizon at $y_0$:
\be
\Delta t= \int ^{y_{final}}_{L_1}  dy/\sqrt{A(y)}
\ee
where $\Delta t$ is clearly divergent when $y_{final}\rightarrow y_0$.
(On the other hand, an observer travelling in the $y$ direction will 
find that he/she takes only finite time to go full circle. See later.) 
So, for an observer on the visible brane, the extra dimension seems 
to have infinite size, since 
a light-like signal he sends out in the positive $y$ direction never 
finish going around the circle to come back to him in finite time. 
In this sense, the world does not look like 4-dimensional to the 
observer on the visible brane, even though the standard model fields, 
being trapped on the visible brane, essentially live in 4 dimensions, 
and gravity behaves like 4-dimensional
due to the warped geometry\cite{RS1}. So quantum corrections 
should be treated in the 5-D theory (or in a full string theory), where 
the quantities $\kappa$, $\sigma_0$, $\sigma_1$, $L_1$ and $L_2$ 
(that appear in the exponentially small $\Le$ (\ref{Lecc}) and the warp factor 
(\ref{warpfactor})) are 5-D renormalized quantities.

This non-4-dimensional feature can be seen in another way.
Although the warp factor traps the massless graviton mode, so ordinary 
gravity behaves like 4-dimensional, this is not the case for the gravity 
KK modes. One may follow the approach of Ref\cite{RS1} of writing the 
graviton mode equation in a conformally flat (or an almost conformally 
flat) metric, where 
we see that the particle horizon is mapped to $\pm \infty$.
So one obtains a continuous KK spectrum, whose wavefunction is spread 
throughout the $z$ co-ordinate, from minus infinity to plus infinity.
In terms of the $y$ co-ordinate, the continuous gravity KK modes have 
their wavefunctions peaked at the particle horizon. 
Even though the extra dimension is compactified, we see that 
the presence of the particle horizon reduces the gravity KK modes to 
a continuous spectrum, which usually implies an uncompactified direction. 
Strictly speaking, one cannot integrate out the extra dimension to 
obtain an effective 4-D theory. If one insists to treat the theory as a
4-D effective theory, one may employ the AdS/CFT 
corespondence\cite{maldacena} to obtain a strongly interacting 
conformal field theory\cite{gubser}. As the 5-D bulk cosmological constant
changes, this conformal field theory changes to another conformal 
field theory, a very unusual situation from the perspective of quantum 
corrections to $\Le$\cite{fjstw}. For example, the 4-D quantum correction
to the Newton's force law is purely a 5-D classical effect \cite{DuffLiu}.

In the uncompactified scenario, we still get 4-D gravity due to the 
trapped graviton mode, $\Delta t$ is again infinite, and the 
gravity KK spectrum is again continuous. (The latter two results follow
more from the presence of particle horizons than from the uncompactified 
property. See Ref\cite{fjstw} for a discussion of the global spacetime 
structure.) The evasion of Weinberg's 
argument is again clear. In summary, although ordinary gravity and the 
standard model of strong and electroweak interactions all look 4-dimensional,
the theory, even as a low energy effective field theory, is intrinsically 
non-4-dimensional in the sense discussed above. This evasion of Weinberg's 
argument is essential for the possibility of solving the cosmological 
constant problem in scenarios of this type. 

Ref\cite{ira} also considers the case where the particle horizon is 
lifted (i.e., the minimum of the warp factor is positive).
This can happen in a number of ways (for example, there is a kinetic 
energy term for a scalar mode). In this scenario, there will appear a 
mass gap in the KK spectrum in the compactified model. As long as the 
mass gap is small enough, the above argument should still apply. In the 
uncompactified scenario, the model remains a 5-D theory.

In models where the radion (brane separation) mode is stabilized, the 
radion mass may be in the electroweak scale and can show up as a giant 
resonance in high energy collisions \cite{goldwise}. 
This radion mode (or other particles) can also appear in the above 
scenario. Fortunately, they will be different from the gravity KK spectrum, 
since the latter will appear in high energy collisions as
a tower of resonances with tensor couplings to matter fields.
Angular distribution measurements can easily distinguish them.

If the Planck brane is our universe (i.e., matter fields live on the 
Planck brane), we can still get an exponentially small cosmological 
constant, and physics is consistent with the radion mode remaining 
unstabilized. In this case, the hierarchy problem must be solved by 
other more conventional approaches, and the coupling of the KK modes 
to matter fields will be too weak to be seen in near future high 
energy colliders. Again, there are no giant resonances with tensor 
couplings in high energy collisions.

\section{Phenomenology}

Let us first review briefly the phenomenology of the giant resonances 
in high energy colliders due to the presence of the gravity KK modes 
in the RS orbifold model\cite{RS2,RS1}. 
Then we shall see that, in the presence of 
particle horizons, the gravity KK spectrum becomes continuous, and 
their couplings to ordinary matter becomes weak. A subtle 
issue about non-zero $H$ that arises will be discussed later.

Here we have in mind that $H$ is very small, and all measurements
involving $G_N$ are at distances much smaller than the cosmic size $1/H$.
Instead of bringing the metric (\ref{metric}) into the conformally flat
form, we can bring the metric (\ref{metric}) into
an almost ``conformally flat'' form \cite{fjstw}, 
\be
\label{acfm}
ds^2= \frac{ H^2}{k^2\sinh^2{\left[ H\left(|z|+z_0\right)\right]}}
[dz^2-dt^2+\exp(2Ht)\delta_{ij}dx^idx^j]
\ee
where the value of $k$ and $z_0 \simeq 1/k$ are in general different on the 
two sides of each brane.
For the region that includes the particle horizon, $z_0$ is defined 
so that when $y=0$ we also have $z=0$.
At the particle horizon, when $y\rightarrow y_0$,
we will have $z\rightarrow \pm \infty$.
We observe that the space between the horizons $\left(-y_0,+y_1\right)$
is mapped into the entire real line in the new $z$ coordinate system.
If the model contains branes separated by particle horizons,
each interval that ends in particle horizons in the $y$ coordinate 
will be mapped into the entire real line in the $z$ coordinate, 
so we end up with a collection of "disconnected" spaces, each space 
containing a subset of branes.
(See Ref\cite{fjstw} for a discussion on the global structure of the 
spacetime.) 

The exponential factor $\exp(2Ht)$ in the metric (\ref{acfm}) 
has no effect on time intervals and distances much 
smaller than the Hubble radius. To get an idea, we see that 
$\Le \approx (10^{-3}eV)^4$ in our universe, so $H \approx 10^{-34} eV$, 
a totally negligible effect in collider physics, where the time scale 
is much smaller than a second. In the limit of $H=0$, we recover the 
conformal metric of the RS model, that is, the particle horizon is 
pushed to $y=\pm \infty$.

Now, let us consider the graviton and the KK modes \cite{RS1} in 
any two brane model where the hierarchy problem is solved with the 
warp factor. 
We are interested in the case where $k/M_{Planck}<1$ and the KK mode 
mass $m \ll k$. The graviton mode is a bound state 
trapped around the Planck (hidden,$\sigma_0$) brane. 
The term in the 4-D effective Lagrangian responsible for the gravitational 
coupling to matter fields on the visible (TeV,$\sigma_1$) brane is given by:
\be
\label{coupleg}
L_{int}=-\frac1{M_{Planck}}T^{\mu\nu}(x) h^{\left(m=0\right)}_{\mu\nu}(x)-
\frac{T^{\mu\nu}(x)}{m_{EW}^{3/2}}
\sum_m \psi _m(z_1) h^{\left(m\right)}_{\mu\nu}(x)
\ee 
where $h^{(m=0)}_{\mu\nu}(x)$ is the usual graviton field and
$h^{(m)}_{\mu\nu}(x)$ is the gravity KK excitation field with mass $m$. 
Here, $T^{\mu\nu}(x)$ is the energy-momentum stress tensor of matter fields 
on the visible brane, 
which is sitting at $z_1$, with canonical kinetic 
terms for the brane fields. Although one starts with typical Planck 
scale masses on both the Planck and the visible brane, 
the masses on the visible brane have absorbed a $\sqrt{A(L_1)}$ factor 
so the typical mass scale on the visible brane now becomes $m_{EW}$. 
Here $\psi _m(z_1)$ is  
the normalized wavefunction of the mass $m$ gravity KK mode at the visible 
brane. Using (\ref{warpfactor}) and the relation between $z$ and $y$, we 
have $k_1z_1= M_{Planck}/m_{EW}$, or
\be
\frac{1}{z_1} \approx  \frac{\sqrt{\sigma_0-\sigma_1}}{M_{Planck}}
\frac{m_{EW}}{M_{Planck}}
\ee
In the RS orbifold model\cite{RS2}, the size of the orbifold is simply 
the brane separation $z_1$. (In Figure 1, the orbifold corresponds to 
the region between $L_0$ and $L_1$ with $\Lambda_1$ in the bulk.)
So the gravity KK modes are discretized, 
with mass $m \simeq n/z_1$, for integer $n$. 
In this model, $\sigma_1 = -\sigma_0$, so the lowest 
KK modes have masses $m =n/z_1 \simeq n\sqrt{\sigma_0}{m_{EW}/M^2_{Planck}}$. 
For relatively large $z_1$, $|\psi_m(z_1)| \simeq 1/\sqrt{z_1}$ 
independent of $m$ for the discrete KK spectrum.
In this case, the effective coupling of the KK modes to the
visible brane matter fields is
\be
L_{KK} \simeq - \frac{\sigma_0^{1/4}}{M_{Planck}}\frac{T^{\mu\nu}(x)}{m_{EW}} 
\sum_{m=n/z_1} h^{\left(m\right)}_{\mu\nu}(x)
\ee
It is reasonable to take the Planck brane tension $\sigma_0$
to be comparable to 
$M_{Planck}^4$, so the lowest resonance is around the electroweak scale
and have electroweak strength coupling to matter fields. 
Since its coupling is stronger than the gravitational coupling by a 
factor of $M_{Planck}/m_{EW}$, giant resonances are expected in high 
energy (TeV scale) collision processes such as $e^+e^-$ annihilations or 
Drell-Yan scatterings \cite{W_Phenomenology}.

Now, let us consider the two brane compactified model (see Figure 1). 
In this scenario, 
the presence of the particle horizon implies that the effective size 
in the $z$ coordinate (conformal metric) is infinite, so the KK spectrum 
is no longer dictated by the position of the visible brane 
(which is still at $z_1$), but by the 
infinite size in the $z$ coordinate. 
The resulting KK spectrum is continuous. 
This implies that there is no distinct resonance signature in high energy 
collisions due to the gravity KK modes.
Since $\sigma_0 > \sigma_1 > 0$, the discussion will be simplified if
we take $\sigma_1$ to be negligibly small. 
The resulting phenomenology is similar to that of a probe brane in the 
presence of a Planck brane in the uncompactified case \cite{lykkenR}. 
For each mass eigenvalue, there are 
a symmetric mode and an anti-symmetric mode (defined to have zero 
wavefunction at the Planck brane), and both sets of
KK modes will contribute at the visible brane. 

For simplicity, consider $\sigma_0 \approx M_{Planck}^4$. 
For KK modes with mass $  H \lesssim  m \lesssim m_{EW}^2/ M_{Planck}$,  
$\psi _m(z_1) \approx -\sqrt{m/{k_1^4z_1^3}}$. 
These modes have similar suppression factor as the trapped gravity mode
and they contribute to the modification of the Newton's law\cite{RS1}. 
For KK modes with mass $ m_{EW} > m > m_{EW}^2/ M_{Planck}$,  
$\psi _m(z_1) \approx (m/m_{EW})^{5/2}$. Using (\ref{coupleg}), we see 
that a typical cross-section behaves like \cite{lykkenR}: 
\be
\sigma \approx E^6/m_{EW}^8
\ee
For energies below $m_{EW}$, the cross-section is quite small.
For energies above $m_{EW}$, we must include KK modes with mass
$ m > m_{EW}$. Their wavefunctions at the 
visible brane is $\psi _m(z_1) \approx 1$, so the scattering involving 
the gravity KK modes becomes strong at energies above $m_{EW}$.
In the $y$ coordinate, this means that the wavefunctions of these KK modes 
are peaked at the particle horizon.
Note that there is no oscillating behavior of $|\psi _m(z)|^2$ 
that is shown in Ref\cite{lykkenR} when both symmetric and 
antisymmetric modes are included.

In processes like $e^+e^- \rightarrow photon + KK mode$, unless the 
KK field bounces back from the Planck brane, it will be lost. Such 
missing energy events may provide a good signal of the scenario when 
the energy approaches $m_{EW}$.

\section{The gravity modes for finite H}

Let us comment on the effect of a non-zero but very small 
$H$ \cite{AdS_Fluctuations,AdS_Fluct_Details,VDVZ_DIscontinuity}.
The setup consists of a Planck brane where all the mass 
scales are comparable to the Planck scale (and consequently, gravity 
is strong), and a visible brane which is treated as a probe brane, 
having little effect on the shape of the wavefunctions of the graviton
along the 5th dimension. Consequently all the jump conditions will 
be imposed at the Planck brane. To solve for the gravity modes, 
we go to the almost conformally flat background metric (\ref{acfm}).
Following Ref\cite{AdS_Fluctuations}, we decompose the fluctuations 
of the metric $ds^2=\left(g_{ab}+\hat h_{ab}\right)dx^adx^b$, into a 
wavefunction along the 5th dimension and a wavefunction in 4D deSitter 
space-time: $\hat h_{\mu\nu}\left(x^{\rho},z\right)
=A\left(z\right)^{-1/4}h_{\mu\nu}\left(x^{\rho}\right)\psi\left(z\right)$.
The equations for $\psi\left(z\right)$ and 
$h_{\mu\nu}\left(x^{\rho}\right)$ are found to be:
\be
\label{massshift}
-\partial_z^2\psi+\frac{\left(A^{3/4}\right)^{\prime\prime}}{A^{3/4}}\psi=
\tilde m^2 \psi, \qquad
-\Box h_{\mu\nu}+\left(2H^2+\tilde m^2\right)h_{\mu\nu}=0
\ee
where $\Box$ indicates the 4D covariant d'Alembertian.
We shall treat the Hubble constant $H$ to be very small.
Following Ref\cite{RS1}, and using the metric (\ref{acfm}),
we obtain the following equation for the gravity modes
\be
\label{Hequation}
-\partial^2_z\psi+\left[\frac{15}{4}H^2\coth^2\left(H\left(|z|+z_0\right)
\right)-\frac{3H^2}{2}
-3H\coth\left(Hz_0\right)\delta\left(z\right)\right]\psi=\tilde m^2\psi 
\ee
This equation has a trapped mode with $\tilde m = 0$ and wavefunction
\be
\psi_0\left(z\right)=\left(\frac{H}{k\sinh\left(H\left(z+z_0\right)\right)}
\right)^{3/2}
\ee
This is the graviton mode. Following (\ref{massshift}), we see that the 
graviton mode has mass $m^2=\tilde m^2+2H^2=2 H^2$. 
We may solve Eq(\ref{Hequation}) in terms of hypergeometric functions
(in the variable $\coth (H(|z|+z_0)$ or $\tanh (H(|z|+z_0)$). 
For our purpose here, 
Since we need only an estimate of the lower bound of 
the continuous KK spectrum, let us consider 
the two regimes: namely, $H|z| \ll 1$ and $H|z| \gg 1$. Since, for the 
visible brane, $z=z_1 \approx m_{EW}^{-1}$, let us first consider the 
regime where $H|z|\ll 1$. 
Using the expansion of the $\coth$ and $\sinh$ functions for small values 
of the argument, we obtain:
\be
-\partial^2_z\psi+\left[\frac{15}{4\left(|z|+z_0\right)^2}-
3\frac{\delta\left(z\right)}{z_0}\left[1+\frac{(Hz_0)^2}{3}\right]
\right]\psi=
\left(\tilde m^2+\frac{3H^2}2\right)\psi=\overline m^2\psi
\ee
which is valid for small $H|z|$. Here we may also drop the term
$(Hz_0)^2/3$ since $Hz_0\ll Hz_1$. 
This reduces to the original Randall-Sundrum equation except that
the mass eigenvalue is replaced by $\overline{m}^2 = \tilde m^2+3H^2/2$.
This yields the following solution:
\be
\label{linecom}
\psi\left(z\right)=
\sqrt{\overline{m}(|z|+z_0)}[aJ_2\left(\overline{m}\left(|z|+z_0\right)\right)+
bY_2\left(\overline{m}\left(|z|+z_0\right)\right)]
\ee
where $\overline{m}^2 = \tilde m^2+ 3H^2/2 \ge 0$.
The solution must also satisfy the junction condition: 
$\psi^{\prime}\left(0+\right)-\psi^{\prime}\left(0-\right)=-3\psi(0)/z_0$.
After properly normalized, we find that \cite{RS1}
\be
 a = 1, \qquad b= \pi\overline{m}^2 z_0^2/4
\ee
so the continuous gravity KK spectrum is bounded below, by $m^2 \ge H^2/2$.
On the visible brane, both symmetric and anti-symmetric modes will be 
present. For large $H|z|$, Eq({\ref{Hequation}) becomes
\be
-\partial^2_z\psi= \left(\tilde m^2- \frac{9}{4}H^2\right)\psi
\ee
which yields either a sinusoidal solution (for $\tilde m^2> 9H^2/4$), 
or an exponential solution (for $\tilde m^2 < 9H^2/4$). 
The linear combination (that is, the coefficients $a$ and $b$) of 
the wavefunction $\psi_m(z)$ in Eq(\ref{linecom})
will change slightly. However, the 
basic physics is unchanged. 
We expect the KK spectrum to start at $m^2 \approx 2H^2$. 
Also the $H \rightarrow 0$ limit is expected 
to be smooth \cite{VDVZ_DIscontinuity}.

Due to the non-zero value of $H$, the Newton's law and the correction 
from the gravity KK modes are changed slightly.
The gravitational potential for 
masses $M_1$ and $M_2$ on the visible brane will be:
\be
V\left(r\right)=-G_N\frac{M_1M_2}{r}\left(e^{-\sqrt{2}Hr}
+ \frac{M_{Planck}^2}{\sigma_0}\int_{\alpha H}^{\infty}
e^{-mr} m dm \right )
\ee
where $\alpha$ is of order 1.
We see that the effects due to the non-zero $H$ are negligible
on distances much smaller than the Hubble radius.

\section{AdS Proper Time}

Earlier, we point out that an observer travelling in the $y$ direction 
will find that it takes only a finite amount of time to go full circle.
To show this, we need to find the proper time for an observer moving along the 
5th dimension towards the particle horizon located at $y=y_0$.
We may use either the Killing vector approach, or equivalently, 
the geodesic approach. Let us follow the latter.  

Since the observer moves along a timelike geodesic, we choose the proper 
time as the affine parameter. Only one connection coefficient, namely 
$\Gamma^y_{tt}$, is non-zero, so the equation of the geodesic becomes:
\be
\label{geodesic_eq}
\frac{d^2y}{d\tau^2}+\Gamma^y_{\mu\nu}\frac{dx^{\mu}}{d\tau}\frac{dx^{\nu}}
{d\tau}=0 \Longrightarrow \frac{d^2y}{d\tau^2}+\frac12\frac{dA\left(y\right)}{dy}
\left(\frac{dt}{d\tau}\right)^2=0
\ee
Using the metric (\ref{metric})
where $A\left(y\right)=H^2\sinh^2\left[k\left(y-y_0\right)\right]/k^2$
and the fact that $x^1,x^2,x^3=constant$ we obtain:
\be
\label{metric_eq}
\left(\frac{dy}{d\tau}\right)^2-A\left(y\right)\left(\frac{dt}{d\tau}
\right)^2=-1
\ee
since in the reference frame moving with the observer, $ds^2=-d\tau^2$.
We can reduce the second-order differential equation to a first-order one 
using the standard method:
\be
\frac{d^2y}{d\tau^2}=\frac{d}{d\tau}\left(\frac{dy}{d\tau}\right)=
\frac{dy}{d\tau}\frac{d}{dy}\left(\frac{dy}{d\tau}\right)=
\frac12\frac{d}{dy}\left(\frac{dy}{d\tau}\right)^2=
\frac12\frac{d\left(y^{\prime}\right)^2}{dy}
\ee
Multiplying Eq.(\ref{geodesic_eq}) by $A\left(y\right)$ and substituting 
$A\left(y\right)\left(dt/d\tau\right)^2$ from Eq.(\ref{metric_eq})
we obtain:
\be
A\left(y\right)\frac{d\left(y^{\prime}\right)^2}{dy}+
\frac{dA\left(y\right)}{dy}\left[1+\left(y^{\prime}\right)^2\right]=0 
\Longrightarrow A\left(y\right)\left[1+\left(y^{\prime}\right)^2\right]
=E^2
\ee
where the constant $E$ is given by the initial energy (velocity) of the 
observer,
so the proper time it takes an observer to reach the particle horizon 
at $y_0$ (where $A(y_0)=0$) is:
\be
\tau=\int_{0}^{y_0}\sqrt{\frac{A\left(y\right)}{E^2-A\left(y\right)}}dy
\ee
which is finite. This means a traveller around the compactified circle 
will find he/she takes only a finite time to go full circle. 
Compared to the observer staying on the brane, this is the ultimate 
twin paradox.

We thank Csaba Cs\'aki, Gia Dvali, \'Eanna Flanagan, Nick Jones and 
Raman Sundrum for discussions.
This research is partially supported by NSF (S.-H.H.T.) and NASA
(I.W.).

\end{document}